%Paper: hep-th/9301058
%From: Jacob Bekenstein <jacob@cosmic.physics.ucsb.edu>
%Date: Thu, 14 Jan 93 17:05:27 -0800

\magnification=900

\def\journal#1#2#3#4#5{#1, {\sl #5\/}{\bf #2}, #3 (19#4)}
\def\prl#1#2#3#4{\journal{#1}{#2}{#3}{#4}{Phys. Rev. Letters }}
\def\mpla#1#2#3#4{\journal{#1}{A#2}{#3}{#4}{Mod. Phys. Letters }}

\def\prd#1#2#3#4{\journal{#1}{#2}{#3}{#4}{Phys. Rev. D }}

\def\npb#1#2#3#4{\journal{#1}{B#2}{#3}{#4}{Nucl. Phys. }}
\def\cmp#1#2#3#4{\journal{#1}{#2}{#3}{#4}{Commun. Math. Phys. }}
\def\ijmpc#1#2#3#4{\journal{#1}{#2}{#3}{#4}{Int. Journ. Mod. Phys. C }}

\def\plb#1#2#3#4{\journal{#1}{#2B}{#3}{#4}{Phys. Lett. }}
\def\rmp#1#2#3#4{\journal{#1}{#2}{#3}{#4}{Revs. Mod. Phys. }}

\def\inbooked#1#2#3#4#5{#1, in {\it #2\/,} ed. #3 ({#4,} 19#5)}

\def\book#1#2#3#4{#1, {\it #2\/} (#3, 19#4)}

\font\twelverm=cmr10 scaled 1200    \font\twelvei=cmmi10 scaled 1200
\font\twelvesy=cmsy10 scaled 1200   \font\twelveex=cmex10 scaled 1200
\font\twelvebf=cmbx10 scaled 1200   \font\twelvesl=cmsl10 scaled 1200
\font\twelvett=cmtt10 scaled 1200   \font\twelveit=cmti10 scaled 1200

\skewchar\twelvei='177   \skewchar\twelvesy='60

%  Define \...point macros to change fonts and spacings consistently

\def\twelvepoint{\normalbaselineskip=12.4pt
  \abovedisplayskip 12.4pt plus 3pt minus 9pt
  \belowdisplayskip 12.4pt plus 3pt minus 9pt
  \abovedisplayshortskip 0pt plus 3pt
  \belowdisplayshortskip 7.2pt plus 3pt minus 4pt
  \smallskipamount=3.6pt plus1.2pt minus1.2pt
  \medskipamount=7.2pt plus2.4pt minus2.4pt
  \bigskipamount=14.4pt plus4.8pt minus4.8pt
  \def\rm{\fam0\twelverm}          \def\it{\fam\itfam\twelveit}%
  \def\sl{\fam\slfam\twelvesl}     \def\bf{\fam\bffam\twelvebf}%
  \def\mit{\fam 1}                 \def\cal{\fam 2}%
  \def\tt{\twelvett}
  \def\nullspace{\nulldelimiterspace=0pt \mathsurround=0pt }
  \def\big##1{{\hbox{$\left##1\vbox to 10.2pt{}\right.\nullspace$}}}
  \def\Big##1{{\hbox{$\left##1\vbox to 13.8pt{}\right.\nullspace$}}}
  \def\bigg##1{{\hbox{$\left##1\vbox to 17.4pt{}\right.\nullspace$}}}
  \def\Bigg##1{{\hbox{$\left##1\vbox to 21.0pt{}\right.\nullspace$}}}
  \textfont0=\twelverm   \scriptfont0=\tenrm   \scriptscriptfont0=\sevenrm
  \textfont1=\twelvei    \scriptfont1=\teni    \scriptscriptfont1=\seveni
  \textfont2=\twelvesy   \scriptfont2=\tensy   \scriptscriptfont2=\sevensy
  \textfont3=\twelveex   \scriptfont3=\twelveex  \scriptscriptfont3=\twelveex
  \textfont\itfam=\twelveit
  \textfont\slfam=\twelvesl
  \textfont\bffam=\twelvebf
  \scriptfont\bffam=\tenbf
  \scriptscriptfont\bffam=\sevenbf
  \normalbaselines\rm}

%tenpoint

%%Various internal macros

\def\beginlinemode{\endmode
  \begingroup\parskip=0pt \obeylines\def\\{\par}\def\endmode{\par\endgroup}}
\def\beginparmode{\endmode
  \begingroup \def\endmode{\par\endgroup}}
\let\endmode=\par
{\obeylines\gdef\
{}}
\def\singlespace{\baselineskip=\normalbaselineskip}
\def\oneandahalfspace{\baselineskip=\normalbaselineskip
  \multiply\baselineskip by 3 \divide\baselineskip by 2}

\def\doublespace{\baselineskip=\normalbaselineskip \multiply\baselineskip by 2}
\newcount\firstpageno
\firstpageno=2
%% FOLLOWING LINE CANNOT BE BROKEN BEFORE 80 CHAR
%% FOLLOWING LINE CANNOT BE BROKEN BEFORE 80 CHAR
\footline={\ifnum\pageno<\firstpageno{\hfil}\else{\hfil\twelverm\folio\hfil}\fi}
\let\rawfootnote=\footnote% We must set the footnote style
\def\footnote#1#2{{\tenrm\singlespace\parindent=0pt\rawfootnote{#1}{#2}}}
\def\raggedcenter{\leftskip=4em plus 12em \rightskip=\leftskip
  \parindent=0pt \parfillskip=0pt \spaceskip=.3333em \xspaceskip=.5em
  \pretolerance=9999 \tolerance=9999
  \hyphenpenalty=9999 \exhyphenpenalty=9999 }
\def\dateline{\rightline{\ifcase\month\or
  January\or February\or March\or April\or May\or June\or
  July\or August\or September\or October\or November\or December\fi
  \space\number\year}}
\def\received{\vskip 3pt plus 0.2fill
 \centerline{\sl (Received\space\ifcase\month\or
  January\or February\or March\or April\or May\or June\or
  July\or August\or September\or October\or November\or December\fi
  \qquad, \number\year)}}

%%Page layout, margins, font and spacing (feel free to change)

\hsize=6.5truein
%\hoffset=1truein
\vsize=8.9truein
%\voffset=1truein
\parskip=\smallskipamount
\twelvepoint% selects twelvepoint fonts (cf. \tenpoint)
\doublespace% selects double spacing for main part of paper (cf.
%\singlespace, \oneandahalfspace)
\overfullrule=0pt % delete the nasty little black boxes for overfull box

%%The user definitions for major parts of a paper (feel free to change)

\def\preprintno#1{\singlespace
 \rightline{\rm #1}}% Preprint number at upper right of title page

\def\title%  Title on title page
  {\null\vskip 3pt plus 0.2fill
   \beginlinemode \doublespace \raggedcenter \bf}

\font\titlefont=cmr10 scaled \magstep3

%\def\author{\vskip 12pt plus 0.2fill \beginlinemode\singlespace \raggedcenter}
%  Author(s) name(s)  on title page

\font\twelvesc=cmcsc10 scaled 1200
\def\author{\vskip 16pt plus 0.2fill \beginlinemode\singlespace
\raggedcenter\twelvesc}
%  same with scs font

\def\affil% Affiliations (can intermix with \author)
  {\vskip 4pt plus 0.1fill \beginlinemode
   \oneandahalfspace \raggedcenter \sl}

\def\abstract  % Begin abstract
  {\vskip 24pt plus 0.3fill \beginparmode
   \narrower\centerline{ABSTRACT}\vskip 12pt }

\def\endpage{\vfill\eject}     %  Eject a page

\def\body{\beginparmode}
 % Begin text body;  can be used to end \title, \author, \affil,
 %  \abstract,\reference, or \figurecaption modes

\def\endtitlepage{\endpage\body}
% End title page, begin body of paper

\def\head#1{% Head;  NOTE enclose the text in {}
  \filbreak\vskip 0.35truein%  e.g., \head{I. Introduction}
  {\immediate\write16{#1}
   \raggedcenter \uppercase{#1}\par}
   \nobreak\vskip 0.2truein\nobreak}

\def\subhead#1{% Subhead;  NOTE enclose the text in {}
  \vskip 0.20truein% e.g., \subhead{A. History of the Problem}
  {\raggedcenter #1 \par}
   \nobreak\vskip 0.15truein\nobreak}

\def\References  %Begin references
  {
   \subhead{References}
   \beginparmode
   \frenchspacing\parindent=0pt \leftskip=1truecm
   \parskip=2pt plus 3pt \everypar{\hangindent=\parindent}}

\gdef\r#1{\indent\hbox to 0pt{\hss#1.~}}   % Ref list numbers e.g., 3.

\gdef\refis#1{\indent\hbox to 0pt{\hss[#1]~}}   % Ref list numbers e.g., [3]

\def\endreferences{\body}

%%    Figures

\def\figurecaptions% Begin figure captions
  {\endpage
   \beginparmode
   \head{Figure Captions}
}

%%   Tables

% To end paper

\def\endpaper{\endmode\vfill\supereject}
\def\endit{\endpaper\end}

%%      Various little user definitions

\def\cite#1{{#1}}
\def\[#1]{[\cite{#1}]}                  % for references like [6]
\def\q#1{\ \[#1]}					                  %        ditto
\def\refto#1{$^{#1}$}                   % For references in text as superscript
\def\ref#1{Ref.~#1}                     %       for inline references
\def\Ref#1{Ref.~#1}                     %       ditto

\def\call#1{{#1}}
\def\(#1){(\call{#1})}
        % For citation of equation numbers
      %       ditto
\def\Eq#1{Eq.~\(#1)}                   %       ditto
                 %       ditto

\catcode`@=11
\newcount\tagnumber\tagnumber=0

\immediate\newwrite\eqnfile
\newif\if@qnfile\@qnfilefalse
\def\write@qn#1{}
\def\writenew@qn#1{}
\def\w@rnwrite#1{\write@qn{#1}\message{#1}}
\def\@rrwrite#1{\write@qn{#1}\errmessage{#1}}

\def\t@ghead{}
\def\taghead#1{\gdef\t@ghead{#1}\global\tagnumber=0}

\expandafter\def\csname @qnnum-3\endcsname
  {{\t@ghead\advance\tagnumber by -3\relax\number\tagnumber}}
\expandafter\def\csname @qnnum-2\endcsname
  {{\t@ghead\advance\tagnumber by -2\relax\number\tagnumber}}
\expandafter\def\csname @qnnum-1\endcsname
  {{\t@ghead\advance\tagnumber by -1\relax\number\tagnumber}}
\expandafter\def\csname @qnnum0\endcsname
  {\t@ghead\number\tagnumber}
\expandafter\def\csname @qnnum+1\endcsname
  {{\t@ghead\advance\tagnumber by 1\relax\number\tagnumber}}
\expandafter\def\csname @qnnum+2\endcsname
  {{\t@ghead\advance\tagnumber by 2\relax\number\tagnumber}}
\expandafter\def\csname @qnnum+3\endcsname
  {{\t@ghead\advance\tagnumber by 3\relax\number\tagnumber}}

\def\equationfile{%
  \@qnfiletrue\immediate\openout\eqnfile=\jobname.eqn%
  \def\write@qn##1{\if@qnfile\immediate\write\eqnfile{##1}\fi}
  \def\writenew@qn##1{\if@qnfile\immediate\write\eqnfile
    {\noexpand\tag{##1} = (\t@ghead\number\tagnumber)}\fi}
}

\def\callall#1{\xdef#1##1{#1{\noexpand\call{##1}}}}
\def\call#1{\each@rg\callr@nge{#1}}

\def\each@rg#1#2{{\let\thecsname=#1\expandafter\first@rg#2,\end,}}
\def\first@rg#1,{\thecsname{#1}\apply@rg}
\def\apply@rg#1,{\ifx\end#1\let\next=\relax%
\else,\thecsname{#1}\let\next=\apply@rg\fi\next}

\def\callr@nge#1{\calldor@nge#1-\end-}
\def\callr@ngeat#1\end-{#1}
\def\calldor@nge#1-#2-{\ifx\end#2\@qneatspace#1 %
  \else\calll@@p{#1}{#2}\callr@ngeat\fi}
\def\calll@@p#1#2{\ifnum#1>#2{\@rrwrite{Equation range #1-#2\space is bad.}
\errhelp{If you call a series of equations by the notation M-N, then M and
N must be integers, and N must be greater than or equal to M.}}\else%
 {\count0=#1\count1=#2\advance\count1
by1\relax\expandafter\@qncall\the\count0,%
  \loop\advance\count0 by1\relax%
    \ifnum\count0<\count1,\expandafter\@qncall\the\count0,%
  \repeat}\fi}

\def\@qneatspace#1#2 {\@qncall#1#2,}
\def\@qncall#1,{\ifunc@lled{#1}{\def\next{#1}\ifx\next\empty\else
  \w@rnwrite{Equation number \noexpand\(>>#1<<) has not been defined yet.}
  >>#1<<\fi}\else\csname @qnnum#1\endcsname\fi}

\let\eqnono=\eqno
\def\eqno(#1){\tag#1}
\def\tag#1$${\eqnono(\displayt@g#1 )$$}

\def\aligntag#1\endaligntag
  $${\gdef\tag##1\\{&(##1 )\cr}\eqalignno{#1\\}$$
  \gdef\tag##1$${\eqnono(\displayt@g##1 )$$}}

\def\eqalignno#1{\displ@y \tabskip\centering
  \halign to\displaywidth{\hfil$\displaystyle{##}$\tabskip\z@skip
    &$\displaystyle{{}##}$\hfil\tabskip\centering
    &\llap{$\displayt@gpar##$}\tabskip\z@skip\crcr
    #1\crcr}}

\def\displayt@gpar(#1){(\displayt@g#1 )}

\def\displayt@g#1 {\rm\ifunc@lled{#1}\global\advance\tagnumber by1
        {\def\next{#1}\ifx\next\empty\else\expandafter
        \xdef\csname @qnnum#1\endcsname{\t@ghead\number\tagnumber}\fi}%
  \writenew@qn{#1}\t@ghead\number\tagnumber\else
        {\edef\next{\t@ghead\number\tagnumber}%
        \expandafter\ifx\csname @qnnum#1\endcsname\next\else
        \w@rnwrite{Equation \noexpand\tag{#1} is a duplicate number.}\fi}%
  \csname @qnnum#1\endcsname\fi}

\def\ifunc@lled#1{\expandafter\ifx\csname @qnnum#1\endcsname\relax}

\let\@qnend=\end\gdef\end{\if@qnfile
\immediate\write16{Equation numbers written on []\jobname.EQN.}\fi\@qnend}

\catcode`@=12

%% DEBUG
%%\def\see#1 {\expandafter\show\csname#1\endcsname}

\catcode`@=11
\newcount\r@fcount \r@fcount=0
\newcount\r@fcurr
\immediate\newwrite\reffile
\newif\ifr@ffile\r@ffilefalse
\def\w@rnwrite#1{\ifr@ffile\immediate\write\reffile{#1}\fi\message{#1}}

\def\writer@f#1>>{}
\def\referencefile{%  Stuff to write .REF file
  \r@ffiletrue\immediate\openout\reffile=\jobname.ref%
  \def\writer@f##1>>{\ifr@ffile\immediate\write\reffile%
    {\noexpand\refis{##1} = \csname r@fnum##1\endcsname = %
     \expandafter\expandafter\expandafter\strip@t\expandafter%
     \meaning\csname r@ftext\csname r@fnum##1\endcsname\endcsname}\fi}%
  \def\strip@t##1>>{}}

\def\citeall#1{\xdef#1##1{#1{\noexpand\cite{##1}}}}
\def\cite#1{\each@rg\citer@nge{#1}}% Variable No. of args, separated by ","

\def\each@rg#1#2{{\let\thecsname=#1\expandafter\first@rg#2,\end,}}
\def\first@rg#1,{\thecsname{#1}\apply@rg}% each@ag is a general purpose
\def\apply@rg#1,{\ifx\end#1\let\next=\relax%  variable no. of arg. macro.
\else,\thecsname{#1}\let\next=\apply@rg\fi\next}% args separated by commas

\def\citer@nge#1{\citedor@nge#1-\end-}% Check for M-N range (M and N numbers)
\def\citer@ngeat#1\end-{#1}
\def\citedor@nge#1-#2-{\ifx\end#2\r@featspace#1 % Single argument
  \else\citel@@p{#1}{#2}\citer@ngeat\fi}% M-N range of arguments
\def\citel@@p#1#2{\ifnum#1>#2{\errmessage{Reference range #1-#2\space is bad.}%
    \errhelp{If you cite a series of references by the notation M-N, then M and
    N must be integers, and N must be greater than or equal to M.}}\else%
 {\count0=#1\count1=#2\advance\count1
by1\relax\expandafter\r@fcite\the\count0,%
  \loop\advance\count0 by1\relax%  Loop from M to N
    \ifnum\count0<\count1,\expandafter\r@fcite\the\count0,%
  \repeat}\fi}

\def\r@featspace#1#2 {\r@fcite#1#2,}% Eat spaces at beginning or end of arg
\def\r@fcite#1,{\ifuncit@d{#1}%  Cite individual reference
    \newr@f{#1}%
    \expandafter\gdef\csname r@ftext\number\r@fcount\endcsname%
                     {\message{Reference #1 to be supplied.}%
                      \writer@f#1>>#1 to be supplied.\par}%
 \fi%
 \csname r@fnum#1\endcsname}
\def\ifuncit@d#1{\expandafter\ifx\csname r@fnum#1\endcsname\relax}%
\def\newr@f#1{\global\advance\r@fcount by1%
    \expandafter\xdef\csname r@fnum#1\endcsname{\number\r@fcount}}

\let\r@fis=\refis% Save old \refis, redefine
\def\refis#1#2#3\par{\ifuncit@d{#1}%      Use two params #2 #3 to strip blank
   \newr@f{#1}%
   \w@rnwrite{Reference #1=\number\r@fcount\space is not cited up to now.}\fi%
  \expandafter\gdef\csname r@ftext\csname r@fnum#1\endcsname\endcsname%
  {\writer@f#1>>#2#3\par}}

\def\ignoreuncited{%   redefine \refis if ignoring uncited references
   \def\refis##1##2##3\par{\ifuncit@d{##1}%
     \else\expandafter\gdef\csname r@ftext\csname
r@fnum##1\endcsname\endcsname%
     {\writer@f##1>>##2##3\par}\fi}}

\def\r@ferr{\endreferences\errmessage{I was expecting to see
\noexpand\endreferences before now;  I have inserted it here.}}
\let\r@ferences=\references
\def\references{\r@ferences\def\endmode{\r@ferr\par\endgroup}}

\let\endr@ferences=\endreferences
\def\endreferences{\r@fcurr=0%  Save old \endreferences, redefine
  {\loop\ifnum\r@fcurr<\r@fcount%  Loop over refnum and produce text
    \advance\r@fcurr by 1\relax\expandafter\r@fis\expandafter{\number\r@fcurr}%
    \csname r@ftext\number\r@fcurr\endcsname%
  \repeat}\gdef\r@ferr{}\endr@ferences}

% Save old \endpaper, redefine it to write parting message.

\let\r@fend=\endpaper\gdef\endpaper{\ifr@ffile
\immediate\write16{Cross References written on []\jobname.REF.}\fi\r@fend}

\catcode`@=12

\citeall\refto% These macros will generate citations
\citeall\ref%
\citeall\Ref%

\def\eg{{\it e.g.,\ }}
\def\ie{{\it i.e.,\ }}

\def\i#1{{\it #1\/}}                          % Italicizes the argument.
\def\spose#1{\hbox to 0pt{#1\hss}}

\def\or{{\rm\ \ or\ \ }}

\def\and{{\rm\ \ and\ \ }}

\def\frac#1#2{{#1 \over #2}}

\def\dex#1{\times 10^{#1}}

\def\Dt{\spose{\raise 1.5ex\hbox{\hskip3pt$\mathchar"201$}}}   % upper
\def\dt{\spose{\raise 1.0ex\hbox{\hskip2pt$\mathchar"201$}}}    % lower

\def\rarrow{\rightarrow}
\def\O{\omega}
\def\bet{{\beta_{\rm bh}}}
\def\beti{{\gamma_{\rm i}}}
\def\Gammi{\Gamma_{\rm i}(\O)}
\def\G0{\Gamma_{\rm i0}}
\def\ei{\varepsilon_{\rm i}(\bet, \O)}
\def\Gam{\Gamma_{\rm sjmp}(\O)}
\def\e{\varepsilon_{\rm sjmp}(\bet, \O)}
\def\si{\sigma_{\rm i}(\bet, \O)}
\def\Sbh{S_{\rm bh}}
\def\beteff{{\beta_{\rm eff}}}
\def\emx{e^{-\bet\hbar\O}}

\def\emy{e^{-\beteff\hbar\O}}
\def\epy{e^{\beteff\hbar\O}}
\def\gfactor{\sum_{\rm i}g_{\rm i}}
\def\gfactorB{\sum_{\rm b}g_{\rm i}}
\def\Imax{\Dt I_{\rm max}}

\null
\preprintno{UCSB-TH-93-02}
\preprintno{January 1993}
\vskip 1.0truein
\oneandahalfspace
\centerline {\titlefont How fast does information leak out from a black hole~?}
\vskip 36pt
\centerline{{Jacob D. Bekenstein}\footnote{*}{\rm E-mail:
jacob@cosmic.physics.ucsb.edu\hfill}}
\vskip 20pt
\centerline{\it Department of Physics, University of California at Santa
Barbara,\/}
\centerline{\it Santa Barbara, CA 93106\/}
\vskip 8 pt
\centerline{{\twelverm and}}
\vskip 8 pt
\centerline{\it The Racah Institute of Physics, Hebrew University of
Jerusalem,\/}
\centerline{{\it Givat Ram, Jerusalem 91904,
Israel\/}\footnote{**}{\rm Permanent address\hfill}}
\vskip 36pt
PACS: 97.60.Lf, 95.30.Tg, 04.60.+n, 05.90.+m
\abstract
\oneandahalfspace

Hawking's radiance, even as computed without account of backreaction, departs
from blackbody form  due to the  mode dependence of the barrier penetration
factor. Thus the radiation is not the maximal entropy radiation for given
energy.  By comparing estimates of the actual entropy emission rate with the
maximal entropy rate for the given power, and using standard ideas from
communication theory, we set an upper bound on the permitted information
outflow rate.  This is several times the rates of black hole entropy decrease
or radiation entropy production.  Thus, if subtle quantum effects not
heretofore accounted for code information in the radiance,  the information
that was thought to be irreparably lost down the black hole may gradually leak
back out from the black hole environs over the full duration of the hole's
evaporation.
\endtitlepage
\oneandahalfspace

Following his theoretical discovery of the black hole radiance that bears his
name, Hawking noted\q{hawk} that such radiation  seems to contradict
accepted quantum physics.  If a black hole forms from matter prepared in a pure
state, and then radiates away its mass in ostensibly thermal radiation, one is
left with a high entropy mixed state of radiation.  This contradicts the
quantum dogma that a pure state will always remain pure under Hamiltonian
evolution.  A related contradiction follows from the
interpretation of black hole entropy as the measure of the information hidden
in the black hole about the ways it might have been formed\q{bekfirst}.
Since fully thermal radiation is incapable of conveying detailed information
about its source, that information remains sequestered as the black hole
radiates, and when it finally evaporates away, the information is lost
forever.  These two contradictions are facets of the black hole information
loss paradox.

Three reactions to the paradox are possible (for reviews see
Refs. \cite{harvstrom} and \cite{presk}). The first is to accept the loss of
information and the trasmutation of pure into mixed state as an inevitable
consequence of the merging of gravity with quantum physics\q{hawk}. Specific
schemes for accomplishing this have been found to be incompatible with locality
or conservation of energy\q{bankssusk}.   A second point of view\q{hawk, nuss}
holds that black hole evaporation  leaves a massive remnant of Planck
dimensions which retains  all the information in question.  This possibility is
not as conservative as it sounds.  According to the bound on specific entropy
or information\q{bekbound}, or considerations from quantum
gravity\q{giddremnant}, an object of Planck mass and dimension can hold only a
few bits of information, so that the posited massive remnants cannot fit the
information bill of a large evaporating black hole. Variations of the remnant
idea, their merits and problems have been discussed in Refs. \cite{giddremnant,
many} and \cite{presk}, among many.  Yet a third view\q{pageinfo, thooft}  is
that exploiting subtle correlations in the radiation, the information manages
to leak back out from the incipient black hole in the course of the
evaporation.  The leak cannot be postponed to the late stages of evaporation
without incurring the problems accompanying remnants\q{nuss, presk}.  For
information leak throughout the evaporation to be a reasonable resolution of
the paradox, it must be shown that  an information flow of the appropriate
magnitude can come out of the black hole's near environs.  A step in this
direction is taken in the present paper.

Lately these three viewpoints have been widely examined by means of the
$1+1$ dimensions dilaton-gravity model of an evaporating black hole proposed by
Callan, Giddings, Harvey and Strominger \q{CGHS}.  This model allows explicit
treatment of the quantum radiance and its backreaction on the hole
(for reviews see \ref{harvstrom}).  Whatever the final outcome of this type of
investigations, it will be nontrivial to project the conclusions from this
model to the  realistic case of $1+3$ black holes. Therefore, any new model
independent aproach  which can address the $1+3$ dimensional case would be of
great conceptual help.  We here employ a thermodynamic argument (which fact
makes it virtually model independent) to show that for the $1+3$ dimensional
Schwarzschild black hole, an outflow of information of the required magnitude
to resolve the information problem is permitted in principle. We do not explore
here specific mechanisms for information extraction, but note that this
subjects  has already  received attention\q{shap}.

Although the Hawking radiance has thermal features, as certified by the
exponential distribution of the number of quanta emitted in each mode, and the
lack of correlations between modes\q{hawkparkwaldbek},  it is not
precisely of blackbody form. The would be blackbody spectrum is distorted by
the
mode dependence of the barrier penetration factor $\Gam$, where $s$ stands for
the particle species, $j$ and $m$ for the angular momentum quantum numbers and
$p$ for the polarization, with $\Gamma<1$ in general\q{hawk}.  For a
Schwarzschild black hole of mass $M$, inverse temperature $\bet=8\pi GM/\hbar$
and entropy $\Sbh$, the average energy in a mode is (henceforth we set
$c=1$)
$$
\e = {\hbar\O\Gam\over e^{\bet\hbar\O}\pm 1}    \eqno(spectrum)
$$
where henceforth the upper (lower) sign corresponds to fermions (bosons).
It is as if blackbody radiation has been passed through a filter.  But the
analogy with filtered radiation stops there.  In the laboratory the filter at
the mouth of a blackbody cavity eventually heats up to the cavity's
temperature,
and so eventually the emerging radiation becomes blackbody.  For a black hole
the distortion is permanent. Perhaps a better analogy is the radiation
from a star which, generally, is far from blackbody because it comes from
layers at different temperatures.

A consequence of the distortion is that, compared with blackbody radiation
with the same power (but, of course, at inverse temperature different
from $\bet$), Hawking radiance is less entropic. Alternatively, for given
inverse temperature, Hawking radiation, in contrast with blackbody radiation,
has free energy, and useful work can be gotten out of it by reshuffling the
energy among the modes.  The entropic defficiency suggests that the radiance
may be carrying information about the state of the quantum fields in the far
past, \ie just the information that is supposed to be lost.  This would, of
course, be impossible if the radiance were exactly blackbody.  In our
stellar analogy, much is learned about a star's atmosphere (composition and
physical conditions) from the departure of its spectrum from blackbody, \eg
spectral lines.

Let us look at the question in the light of quantum communication
theory (for  reviews  see \Ref{review}). We shall adapt Lebedev and Levitin's
pioneering thermodynamic approach\q{leblev}, and measure information in
natural units (nits); 1 nit $= \log_2 e $ bits.   To this end we consider the
entropy of the Hawking radiance as entropy (uncertainty about the state) of the
noise which is adulterating the signal conveying the information. The radiance
power, $\Dt E$,  will be interpreted as the sum of noise and signal powers.
With this scenario the maximum rate at which information can be recovered from
the radiation by a suitable detector is $\Imax\equiv\Dt S'-\Dt S$ where $\Dt S$
is the actual entropy outflow rate, while $\Dt S'$ is the maximum entropy rate
corresponding to the actual power $\Dt E$ under the boundary conditions of the
system.  (Actually, if the noise is correlated with the signal, as may well be
the case in the Hawking radiance, $\Imax$ will be larger\q{review}; in this
case our arguments below are actually strengthened).  Lebedev and Levitin
considered a one dimensional communication channel.  Most of our discussion
will be devoted to the issue of how to define the three-dimensional channel
issuing from a black hole.

For convenience we shall use the notation ${\rm i}\equiv\{{\rm sjmp}\}$.  The
probability distribution for the black hole to spontaneously emit $n$ quanta in
mode $\{{\rm i}, \O\}$ is given by\q{hawkparkwaldbek}
$$
p_{\rm sp}(n)=(1\pm e^{-\beti})^{\mp 1}e^{-\beti n}
\eqno(probability)
$$
where  $\beti(\bet, \O)$ is defined by
$$
{1\over e^\beti\pm 1} = {\Gammi\over e^{\bet\hbar\O}\pm 1}      \eqno(defbeti)
$$
{}From this follows the entropy in the given mode:
$$
\si=\pm\ln (1\pm e^{-\beti}) + {\beti\over e^\beti\pm 1}
\eqno(modeentropy)
$$
We may also reexpress \Eq{spectrum} as
$$
\ei= {\hbar\O\over e^\beti\pm 1}      \eqno(modeenergy)
$$
The entropy outflux rate and the power may now be expressed as
$$
\Dt S=\sum_{\rm i}\int^\infty_0 \si{d\O\over 2\pi}    \eqno(Srate)
$$
$$
\Dt E=\sum_{\rm i}\int^\infty_0\ei{d\O\over 2\pi} \eqno(Erate)
$$
where $d\O/2\pi$ is the rate at which modes of type i emanate from the
hole.

Page\q{page1, page2} has calculated numerically the contributions of various
particle species to $\Dt E$ and $\Dt S$, and states the results in terms of
the dimensionless ratios  $\mu\equiv\Dt E(GM)^2\hbar^{-1}$ and $\nu\equiv\Dt
S/(\bet\Dt E)$. For each species of light neutrinos or antineutrinos he finds
$\mu=4.090\dex{-5}$ and $\nu=1.639$ with modes having $j={1\over 2}, {3\over
2}, {5\over 2}$ being the overwhelming contributors.  For photons
$\mu=3.371\dex{-5}$ and $\nu=1.500$ with modes having $j=1,2,3$ making the
dominant contribution.  And for gravitons  $\mu=3.84\dex{-6}$ and
$\nu=1.348$ with modes having $j=2,3$ contributing overwhemingly.  If  the
black hole emits three species of neutrinos and three of antineutrinos (each
with a single helicity), photons and gravitons, the  overall numbers are
$\mu=2.829\dex{-4}$ and $\nu=1.619$ (this last value involves the individual
$\nu$'s weighted by the $\mu$'s). If there are only two light neutrino species
the numbers are $\mu=2.011\dex{-4}$ and $\nu=1.610$.  The contributions of
bosons alone are  $\mu=3.755\dex{-5}$ and $\nu=1.484$.

We now have to compare $\Dt S$ with the entropy rate $\Dt S'$ of the maximally
entropic (blackbody) distribution whose power $\Dt E'$ equals $\Dt E$.   A
straightforward way to get $\Dt S'$ is to compute both it and  $\Dt E'$ from
the Boltzmann formulae for blackbody emission by assuming some effective
radiating area for the black hole.  This ``photosphere'' is not a well defined
concept, depending as it does on frequency.  Thus, in the scattering of high
frequency (geodesically moving) quanta by our black hole, all quanta hitting
within a crossection $27\pi G^2 M^2$ will be captured\q{MTW}.  This suggests a
photospheric area $A_{\rm phot}$ four times as large.  But  $A_{\rm phot}$ must
be larger than that. This is because each $\Gammi$ vanishes only as some power
of $\O$ as $\O\rarrow 0$, so that quanta with fairly large impact parameter are
sometimes absorbed and, therefore, must also be emitted sometimes.  The modes
they are in must thus be included in calculating the comparison blackbody
radiation.  We thus write  $A_{\rm phot}=\xi 108\pi G^2 M^2$ with $\xi > 1$.

At inverse temperature $\beteff$, a black body of area $A_{\rm phot}$
emits power $\Dt E'= N\pi^2 A_{\rm phot}/(60\,\beteff^4\hbar^3)$, where $N$
is the effective number of particle species emitted. Photons and gravitons
contribute $1$ each to $N$; each species of fermions contributes $7/16$.
Therefore, we must set $N=37/8$ if there are three light neutrino species and
$N=15/4$ if there are two.  Comparing these results for $\Dt E'$ with Page's
results for $\Dt E$ we obtain  $\beteff=1.230\,\xi^{1/4}\bet$ for three light
neutrinos and  $\beteff=1.271\,\xi^{1/4}\bet$ for two.  For blackbody radiation
flowing in three space dimensions, $\Dt S'=\frac43\beteff\Dt E'$. Thus, taking
into account the $\nu$ factors of Page,
$$
\Imax\equiv\Dt S'-\Dt S= (1.640\,\xi^{1/4}-1.619)\,\bet\Dt E   \eqno(info)
$$
for three neutrino species; for two the numerical factor is
$1.694\,\xi^{1/4}-1.610$.  Since $\bet\Dt E=-\bet\Dt M=-\Dt \Sbh$ we see that
even if $\xi=1$, 1-5\% of the sequestered information could come out
in principle.  This is just a lower bound because we actually
expect $\xi$ to be larger, so that $\Dt\Imax$ may actually be a substantial
fraction of $|\Dt\Sbh|$.  The present method is, however, unable to tell us
just how much.

The following alternative approach is able to set an {\it upper\/} bound on
$\Dt\Imax/|\Dt\Sbh|$.  We shall compare Page's $\Dt S$ with the entropy flow
$\Dt S'$ in the {\it same\/} angular momentum modes of a blackbody
distribution with inverse temperature $\beteff$  determined by the
equality of the powers (in the relevant set of modes only).  This is
different from the above calculation which compared with blackbody modes
having sharply defined directions (coming from the black hole).  The new
comparison should overestimate $\Dt S'$ and consequently $\Dt\Imax$ because
blackbody radiation populating a finite number of angular momentum modes
assigns
substantial weight to modes with $\O\rarrow 0$; for nonzero orbital angular
momentum these correspond to arbitrarily large impact parameter, and are
thus not related to the black hole.  These spurious modes broaden the phase
space and so artificially increase the entropy rate $\Dt S'$.  Later on we
shall show how to repair part of this problem.

To calculate the blackbody quantities we replace for each mode
$\beti\rarrow\beteff\hbar\O$ in \Eq{modeentropy} and
\Eq{modeenergy}  . The integral of  the logarithmic term in $\sigma_{\rm i}$
of \Eq{Srate}  can be combined with the other term by integration by parts.
Using
$$
\int^\infty_0{{x\,dx\over e^x\pm 1}}={\pi^2(3\mp 1)\over 24}
\eqno(integral)
$$
one can cast the results in the form
$$
{\Dt S'\over 2\beteff}=\Dt E'={\pi\over12\hbar\beteff^2}\gfactor
\eqno(primerates)
$$
where $g_{\rm i}=1\or {1\over 2}$ for a boson or fermion mode, respectively.
Summing over the particle species and modes which Page considered, we obtain
$\gfactor = 90$ for three light neutrinos and $\gfactor = 78$ for two.
Equating
$\Dt E'$ with  Page's $\Dt E$  gives $\beteff=11.48\,\bet$ and
$\Dt S'= 22.97\,\bet\Dt E$ if there are three light neutrinos while
$\beteff=12.68\,\bet$  and  $\Dt S' = 25.36\,\bet\Dt E$ if there are two.
Using the cited values of $\nu$  we conclude that
$$
\Imax = \Dt S'-\Dt S= 21.35\,|\Dt\Sbh|      \eqno(info1)
$$
for three light neutrinos. If there are only two, the numerical factor is
$23.75$.

Although the above figure for  $\Dt\Imax$ is an overestimate, it is so large
as to suggest that an information leak of suficient magnitude to resolve the
information problem is allowed.  For example, if $\Dt I$, the actual
information
outflow rate, amounts to $1.619|\Dt\Sbh|$ throughout the course of evaporation
of a massive black hole down to $M\approx 1\dex{14}\,{\rm g}$ (when the
emission of massive particles becomes important and most of the initial black
hole entropy has disappeared\q{page1}), the outgoing information equals the
total Hawking radiance entropy.  Hence, given an appropriate mechanism, the
radiation can end up in a pure state.

But how exagerated is the above bound on $\Dt\Imax$~?  We shall not attempt
to exclude the low frequency modes by hand from our calculation; such a
task would be fraught with ambiguities.  Rather we ask, \i{if} it were possible
to modify the curvature barrier surrounding the black hole, and consequently
to modify the $\Gammi$, what would be the most entropic spectrum that could
come out of the black hole~?  As we shall see presently, the answer is not
blackbody: the $\Gammi$ cannot all be unity.  However, the new spectrum is
more relevant for comparison than the pure blackbody one because, for given
angular momentum, it does supress low frequency modes.

If we could manipulate the $\Gammi$, the largest entropy flow $\Dt S'$ would
be obtained with the $\Gammi$ as large as possible.  This is seen by
differentiating $\sigma_{\rm i}$ [\Eq{modeentropy}] with respect to $\beti$,
and  transforming the derivative to one with respect to the corresponding
$\Gammi$ with  help of \Eq{defbeti}; the result is positive definite.  We are
thus interested in the hypothetical situation when all the $\Gammi$ are as
large
as physically possible.  For fermion modes no reason is known to prevent
$\Gammi$ from approaching unity.  However, for
boson modes the value of $\Gammi$ is subject to a bound.

This bound stems from the formula
$$
\Gammi= (1-\emx)\,\G0(\O) \eqno(formula)
$$
where $1-\G0$ is the probability that a single incident quantum is scattered
back from the  black hole\q{bekmeis, bekne'eman}. Formula (\call{formula})
follows by combinatorics from the interpretation in terms of a combination of
scattering, and spontaneous and stimulated emission of the conditional
probability  $p(m|n)$  that the Schwarzschild black hole returns outward $m$
quanta in a mode which had $n$ incident ones.  The $p(m|n)$ has been obtained
independently by information theoretic\q{bekmeis} and field
theoretic\q{waldpan} methods.  It turns out to be impossible to understand its
form as due to a combination of Hawking emission and scattering\q{bekmeis}; the
inclusion of stimulated emission supplies the missing element.  The stimulated
emission depresses the value of $\Gammi$ under the naive
absorption probability $\G0(\O)$.  In fact, because $\G0\leq 1$, $\Gammi\leq
1-\emx$.  (The case $\Gamma_{\rm i}=\G0=1$ is not actually excluded by the
considerations of \Ref{bekmeis}, but a $\Gamma_{\rm i}$ close to unity is not
allowed).

In view of the above, let us compare the actual $\Dt S$ with  the
$\Dt S'$ of a spectrum of the form of \Eq{spectrum} with inverse
temperature $\beteff$ and having  $\Gammi=1$ (perfect blackbody) for all
fermion
modes, but $\Gammi=1-\emy$ for all boson modes.  This is the closest a black
hole emission spectrum could come to blackbody, and thus gives the largest
$\Dt S'$ for given power.   Note that the new comparison spectrum is
poor in low frequency bosons as compared with the blackbody spectrum.  Thus we
have gone part of the way towards repairing the problem noted earlier. Since
the fermions are blackbody as in  our previous calculation, we shall just
concentrate on the boson contributions to $\Dt S$ and $\Dt S'$.  In what
follows the subscript ``b"  stands for bosons.

We shall first compute the boson contribution $\Dt E_{\rm b}'$. From the chosen
$\Gammi$ it follows that $\varepsilon_{\rm i}(\beteff,\O)=\hbar\O\emy$.
Thus \Eq{Erate} gives $\Dt E'_{\rm b}=(2\pi\hbar\beteff^2)^{-1}\gfactorB$.
The sum over the boson modes calculated by Page is $54$. Equating the
result to his $\Dt E_{\rm b}$ determines that $\beteff=19.04\,\bet$. We now
compare $\varepsilon_{\rm i}$ with \Eq{modeenergy} to determine that
$e^\beti=1+\epy$.  It then follows from \Eq{modeentropy} that $\sigma_{\rm
i}(\beteff,\O)=\ln(1+\emy)+\emy\ln(1+\epy)$ for boson modes.  After integration
by parts, \Eq{Srate} gives
$$
\Dt S'_{\rm b}=\left(\pi/24+\ln 2/\pi\right){\gfactorB\over\hbar\beteff}
\eqno(last)
$$
Comparing with Page's
result for $\Dt S_{\rm b}$ gives
$$
(\Imax)_{\rm b} = \Dt S'_{\rm b}-\Dt S_{\rm b}= 5.382\,|\Dt\Sbh|   \eqno(info2)
$$
where we have used the value of the \i{total} $\Dt\Sbh$ including the
contribution of three light neutrino species.

Although the above figure may overestimate $(\Imax)_{\rm b}$, we
must still add a contribution from fermions to get the total $\Imax$.
Thus our earlier impression from \Eq{info1} that the departure of
Hawking radiance from blackbody is enough to permit a large information
outflux stands.  Gradual escape of the
sequestered information (equal to $\Sbh$) and reconstitution of a pure
radiation state by the time the hole has evaporated away seem feasible,
provided some quantum mechanism  codes the information in the
radiation.  It would be surprising if nature has not taken advantage of this
opportunity to obviate the information problem.  There remains the task of
identifying the mechanism of information leak. The prominent part played by the
curvature barrier in deforming the blackbody spectrum makes processes
associated with it, such as stimulated emission,  likely culprits.

\vskip 16pt

I thank Gary Horowits, Don Page and Andy Strominger for informative
conversations, and Jim Hartle for hospitality in Santa Barbara.

\vskip 36pt
\References
\oneandahalfspace
\refis{bekmeis}\prd{J. D. Bekenstein and A. Meisels}{15}{2775}{77}.

\refis{bekfirst}\prd{J. D. Bekenstein}{7}{2333}{73}.

\refis{hawkparkwaldbek}\prd{L. Parker}{12}{1519}{75};
\prd{J. D. Bekenstein}{12}{3077}{75}
\cmp{R. M. Wald}{45}{9}{75};
\prd{S. W. Hawking}{13}{191}{76}.

\refis{hawk}\prd{S. W. Hawking}{14}{2460}{76} and {\sl Commun. Math.
Phys.\/ }{\bf 87}, 395 (1982).

\refis{many}\npb{L. Susskind and L. Thorlacius}{382}{123}{92};
\prd{T. Banks, A. Dabholkar, M.R. Douglas, and M. O'Loughlin}{45}{3607}{92};
{T. Banks, A. Strominger and M. O'Loughlin, ``Black hole
remnants and the information puzzle", RU-92-40 and hep-th/9211030}.

\refis{nuss}\plb{Y. Aharonov, A. Casher and S. Nussinov}{191}{51}{87}.

\refis{bekbound}\prd{J. D. Bekenstein}{23}{287}{81}; \prd{J. D. Bekenstein and
M. Schiffer}{39}{1109}{89}.

\refis{bankssusk}\npb{T. Banks, M. E. Peskin and L. Susskind}{244}{125}{84}.

\refis{giddremnant}\prd{S. Giddings}{46}{1347}{92}.

\refis{page1}\prd{D. N. Page}{13}{198}{76}.

\refis{page2}\prd{D. N. Page}{14}{3260}{76}.

\refis{bekne'eman}\inbooked{J. D. Bekenstein}{To Fulfill a Vision}{Y.
Ne'eman}{Addison-Wesley, Reading, Mass.}{81}.

\refis{CGHS}\prd{C.G. Callan, S.B. Giddings, J.A. Harvey, and A.
Strominger}{45}{R1005}{92}.

\refis{pageinfo}\prl{D. Page}{44}{301}{80}.

\refis{thooft}\npb{G.'t Hooft}{256}{727}{85} and {\bf B335}, 138
(1990).

\refis{waldpan}\prd{P. Panangaden and R. M. Wald}{16}{929}{77}.

\refis{MTW}\book{C. W. Misner, K. S. Thorne and J. A.
Wheeler}{Gravitation}{Freeman, San Francisco}{73}.

\refis{review}\rmp{Y. Yamamoto and H. A. Haus}{58}{1001}{86};
\ijmpc{J. D. Bekenstein and M. Schiffer}{1}{355}{90}.

\refis{leblev}\journal{D. S. Lebedev and L. B. Levitin}{149} {1299}{63}{Dokl.
Akad. Nauk SSSR } \enskip [{\sl Sov. Phys. Dokl.\/ }{\bf 8}, 377 (1963)].

\refis{harvstrom}J. Harvey and A. Strominger, ``Quantum aspects of black
holes'', Chicago preprint EFI-92-41,  hep-th/9209055;
S. Giddings, ``Toy models for black hole evaporation'',
UCSB-TH-92-36, hep-th/9209113.

\refis{presk}J. Preskill, ``Do black holes destroy information?''
CALT-68-1819, hep-th/9209058.

\refis{shap}\mpla{J. Preskill, P. Schwarz, A. Shapere, S. Trivedi and F.
Wilczek}{6}{2353}{91}.

\endreferences

\endit\end